\documentclass[prl,twocolumn,preprintnumbers,superscriptaddress]{revtex4}

\usepackage{latexsym} 
\usepackage{graphicx}
\usepackage{dcolumn}
\usepackage{bm}

\begin{document}


\title{Field-induced carrier delocalization in the strain-induced Mott insulating state of an organic superconductor}

\author{Yoshitaka Kawasugi} \email{kawasugi@riken.jp}
 \affiliation{Saitama University, Saitama, Saitama 338-8570, Japan}
 \affiliation{RIKEN, Hirosawa, Wako, Saitama 351-0198, Japan}
\author{Hiroshi M. Yamamoto} \email{yhiroshi@riken.jp}
 \affiliation{RIKEN, Hirosawa, Wako, Saitama 351-0198, Japan}
\author{Naoya Tajima}
 \affiliation{RIKEN, Hirosawa, Wako, Saitama 351-0198, Japan}
\author{Takeo Fukunaga}
 \affiliation{RIKEN, Hirosawa, Wako, Saitama 351-0198, Japan}
\author{Kazuhito Tsukagoshi}
 \affiliation{MANA, NIMS, Tsukuba, Ibaraki 305-0044, Japan}
\author{Reizo Kato}
 \affiliation{Saitama University, Saitama, Saitama 338-8570, Japan}
 \affiliation{RIKEN, Hirosawa, Wako, Saitama 351-0198, Japan}
 

\begin{abstract}
We report the influence of the field effect on the dc resistance and Hall coefficient in the strain-induced Mott insulating state of an organic superconductor $\kappa$-(BEDT-TTF)$_{2}$Cu[N(CN)$_{2}$]Br.
Conductivity obeys the formula for activated transport $\sigma _{\Box} = \sigma _{0}\exp(-W/k_{B}T)$, where $\sigma_{0}$ is a constant and $W$ depends on the gate voltage.
The gate voltage dependence of the Hall coefficient shows that, unlike in conventional FETs, the effective mobility of dense $hole$ carriers ($\sim1.6\times 10^{14}$ cm$^{-2}$) is enhanced by a $positive$ gate voltage.
This implies that carrier doping involves delocalization of intrinsic carriers that were initially localized due to electron correlation.
\end{abstract}


\maketitle
Electrostatic carrier doping (ESD) into novel materials using the field-effect transistor (FET) principle has recently attracted considerable interest \cite{Ahn}. 
One reason for this interest is that ESD could be used to achieve continuous carrier density tuning in materials with minimal disturbance of their underlying lattices \cite{Inoue}.
One fascinating material for ESD is Mott insulators associated with the effect of chemical doping by which unconventional superconductivity can be achieved \cite{Mannhart, Triscone}.

Recently, we reported that a thin-film single crystal of an organic superconductor $\kappa$-(BEDT-TTF)$_{2}$Cu[N(CN)$_{2}$]Br (BEDT-TTF = bis(ethylenedithio)tetrathiafulvalene, abbreviated as $\kappa$-Br) \cite{Williams,Ichimura} adhered to a Si substrate becomes a Mott insulator due to the negative pressure exerted through the incompressible Si substrate. 
In addition, we found that the insulating state exhibits a field effect in which the conductivity enhancement exceeds seven orders of magnitude \cite{Kawasugi}.
The series $\kappa$-(BEDT-TTF)$_{2}$X (X: monoanion) consists of conducting layers of BEDT-TTF and insulating anion layers.
It belongs to the mother system of unconventional superconductivity and have been investigated in terms of the bandwidth-controlled Mott transition \cite{Kanoda}.
In addition, this material group also contains a doped superconductor \cite{Lyubovskaya,Taniguchi}.
On the basis of the generic bandwidth-bandfilling phase diagrams \cite{Imada, Moriya, Watanabe}, the ground state of strained $\kappa$-Br is considered to be located near the tip of the Mott insulating phase.
Consequently, the field effect in a $\kappa$-Br FET is a result of carrier doping in the Mott insulating state, in which dense carriers are present but are localized due to the strong electron correlation.

\begin{figure}[htbp]
 \begin{center}
 \includegraphics{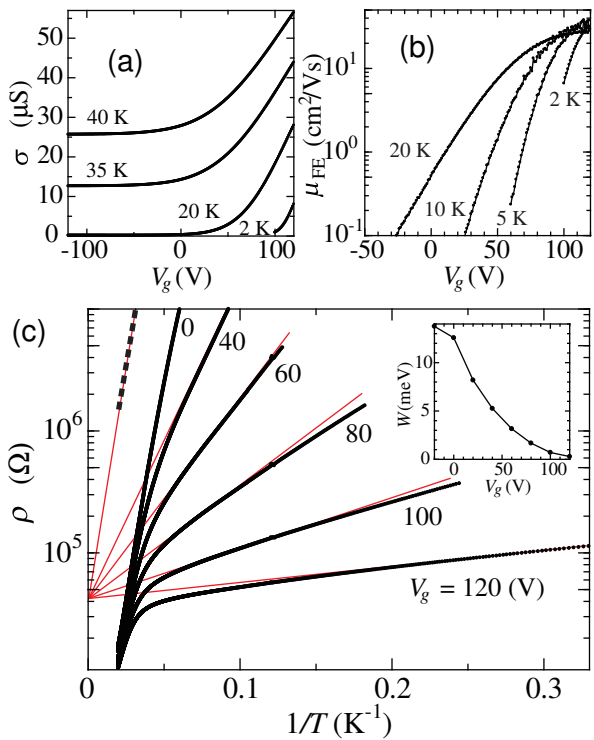}
 \end{center}
 \caption{(a) The sheet conductance variation through gate voltage cycles between $\pm$120 V, at 40, 35, 20, and 2 K. Both the forward and backward data are shown. (b) The field-effect mobilities plotted on a logarithmic scale. (c) Arrhenius plots for the sheet resistance for various gate voltages. The solid lines represent the data for $V_{g}$=120, 100, 80, 60, 40, 0 V in ascending order. The dashed line shows the data for 0 V multiplied by the number of conducting layers. The $V_{g}$ dependence of $W$ is shown in the inset.} 
 \label{fig:Fig1.eps}
\end{figure}

In typical FET devices, the field effect originates from the control of the Fermi energy in the rigid band structure.
By contrast, in a Mott insulator, doped carriers increase screening of Coulomb interaction, resulting in reduction of the effective Coulomb repulsion energy \cite{Kotliar,Mona}.
The energy dependence of the density of state is hence coordinated with the doping concentration so that the charge gap collapses as the bandfilling deviates from 1/2.
In the first part of this letter, we discuss influence of the field effect on the conductivity.
We then discuss the gate voltage dependence of the Hall effect in the second part of this letter.

We constructed a bottom-gate FET configuration by laminating electrochemically synthesized thin single crystals of $\kappa$-Br onto a $p$-doped Si substrate (used as the gate electrode) with a 200 nm thick SiO$_{2}$ layer on which Au source-drain electrodes were evaporated.
The crystal transfer was carried out in ethanol so that the crystal was fixed when the ethanol was evaporated.
Subsequently, the crystal was shaped into a Hall bar geometry using pulsed laser radiation \cite{Yagi}.
Details about sample fabrication and the strain effect have been described in a previous report by us \cite{Kawasugi}.
We employed a conventional five-probe dc method using the standard Hall bar configuration with a current of 1 $\mu$A.
The temperature and the magnetic field normal to the films were controlled using a Quantum Design Physical Property Measurement System.
The unlabeled data are from the same sample (sample A), which had dimensions of 300 $\mu$m$\times$100 $\mu$m$\times$130 nm (width, length, thickness).
For the Hall measurements, data from another sample (sample B), which had dimensions 100 $\mu$m$\times$50 $\mu$m$\times$170 nm), are also shown.

Figure 1 shows the field effect on the conductivity.
The variation in the four-probe conductivity with gate voltage $V_{g}$ (Fig. 1(a)) is reversible and the field-effect mobility $\mu _{\rm FE}\equiv (1/C_{i})(\partial \sigma _{\Box}/\partial V_{g})$ ($C_{i}$: capacitance per unit area of the SiO$_{2}$ film) is around 30 cm$^{2}$/Vs in the linear region for this sample (Fig. 1(b)).
The value of $\mu _{\rm FE}$ at the highest gate voltage increases on cooling to 2 K.
The channel conduction is almost ohmic in the range applied here and exhibits no noticeable hysteresis.
The field effect is reproducible, although the threshold voltage $V_{th}$ is sample dependent.
We observed the ambipolar field effect in samples with relatively high $V_{th}$.
This indicates that the present device is ambipolar, as expected for Mott insulators.
However, the threshold shifts depend on the fabrication process used and the surface conditions of the samples. 
Most of the samples exhibit $n$-type behavior.

Figure 1(c) shows the four-probe sheet resistance in activation plots for different gate voltages.
The activation plot at $V_{g} = 0$ V gives a charge gap of 25 meV for the ungated film (i.e., bulk), whereas the activation energy monotonically decreases with an increase in the gate voltage. 
The gated resistance at low temperature can also be fitted to the activation plots. 
The bends at the intermediate temperature are therefore attributed to the shifts from the bulk conductance regime (high temperature) to the field-induced conductance regime (low temperature).
At $V_{g}$ = 120 V, the activation energy is reduced to about 0.3 meV (3.5 K).
Extrapolations of the field-induced conductance intersect at the high temperature limit at which $\sigma _{0}=2.4\times10^{-5}$ S $\sim$ 0.1 $e^{2}/\hbar$ for this sample.
One finds,
\begin{equation}
\sigma _{\Box} = \sigma _{0}e^{-W/k_{B}T}
\label{eq1}
\end{equation}
where $\sigma _{0}$ is a constant and $W$ is the activation energy, which depends on the gate voltage.
Thus, the gated transport is basically an activation type. 
In FETs with a high density of deep trap states, the activation plots of the conductivity intersect at a finite temperature so that $\sigma _{0}$ decreases with increasing $V_{g}$ (the Meyer-Neldel rule \cite{Meyer}).
On the other hand, in Si-MOSFETs with long-range potential fluctuations, $\sigma _{0}$ increases as $V_{g}$ is increased due to their macroscopic inhomogeneity \cite{Arnold}.
Therefore, $\sigma _{0}$ in the present system indicates that neither deep trap states nor macroscopic inhomogeneity govern the transport properties.
At the high temperature limit, the bulk conductivity ($V_{g}$ = 0 V) divided by the number of conducting layers is almost $\sigma _{0}$, suggesting that the field-induced conducting layer is confined within the half of the unit cell containing a pair of BEDT-TTF and anion layers (1.5 nm).

Let us consider the mechanism for the change in the activation energy in the present system.
In typical FET devices such as Si-MOSFETs, the insulating state is understood in terms of strong Anderson localization \cite{Ando} where $W=E_{C}-E_{F}$ ($E_{C}$: mobility edge, $E_{F}$: Fermi energy).
Application of a gate voltage changes $E_{F}$, resulting in the variation in $W$.
In most of FET devices including this type, the gate voltage does not alter the energy dependence of the density of states (rigid band model).
On the other hand, a sufficient shift of the chemical potential (bandfilling) reduces the effective Coulomb repulsion energy in a Mott insulator.
Realistic theories and experiments predict that the shift in the chemical potential of a Mott insulator modifies the profile of the density of states so that the charge gap collapses \cite{Imada}.
We investigated the Hall effect in the present system to ascertain the nature of the field-induced carriers.
The conductivity can be separated into two components: the carrier density and mobility.

In general, the Hall coefficient is related to the carrier density by $R_{H}=1/en$ ($R_{H}$: Hall coefficient $n$: carrier density).
The electron density is expected to increase linearly with the gate voltage in an ideal $n$-type FET.
We employed two methods, magnetic-field sweeping ($B$-sweeping) and gate-voltage sweeping ($V_{g}$-sweeping) at fixed temperatures.
In $B$-sweeping, the magnetic field varied between $\pm$8.5 T by 100 G/s at fixed gate voltages.
In $V_{g}$-sweeping, the gate voltage cycled between $\pm$120 V under stationary magnetic fields of $\pm$8.5 T. 
The forward and backward data in the reciprocation cycles were confirmed to be consistent for both methods.

\begin{figure}[htbp]
 \begin{center}
 \includegraphics{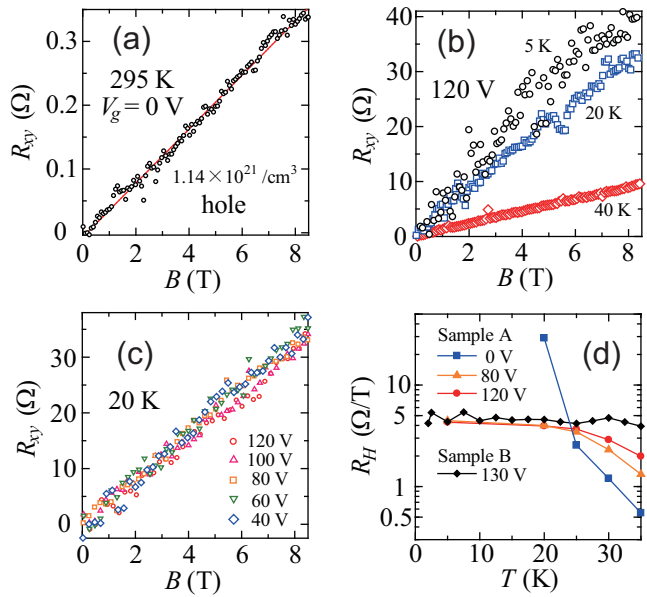}
 \end{center}
 \caption{(a) The Hall resistance vs magnetic field at room temperature, (b) for $V_{g}$=120 V at 40, 20, and 5 K, (c) at 20 K for different gate voltages. (d) The temperature dependence of $R_{H}$.}
 \label{fig:Fig2.eps}
\end{figure}

First, the Hall effect without a gate voltage was measured to confirm the property of the bulk (about 90 conducting BEDT-TTF layers).
$\kappa$-Br has a single folded Fermi surface giving a hole density of $1/eR_{H}$, which corresponds to almost 100\% of the first Brillouin zone at high temperatures.
At room temperature, we obtained a hole density of $n=1.14\times 10^{21}$ cm$^{-3}$ (Fig. 2(a)), which corresponds to 95\% of the active hole density estimated from the lattice parameters and the Hall mobility $\mu _{H}\equiv R_{H}\sigma=0.12$ cm$^{2}$/Vs.
The values obtained are consistent with the reported values for a related material \cite{Murata} because the strain effect is absent at room temperature.
The Hall mobility increased slightly on cooling to about 150 K, after which it started to decrease moderately as the temperature was further reduced.
The carrier density and Hall mobility at 20 K are $n=1.7\times10^{18}$ cm$^{-3}$ and $\mu _{H}=0.09$ cm$^{2}$/Vs.
Therefore, the temperature dependence of the bulk conductivity is almost entirely due to the variation in the carrier density down to intermediate temperatures.

The Hall effect when a gate voltage is applied is found to be anomalous.
The Hall resistances at $V_{g}=120$ V are plotted in Fig. 2(b).
Surprisingly, the Hall coefficients are clearly positive despite the $positive$ applied gate voltage.
Using $n=1/eR_{H}$, we obtain a hole density of $1.6\times 10^{14}$ cm$^{-2}$, which exceeds the surface electron density of $7\times 10^{12}$ cm$^{-2}$ estimated from a typical capacitance model $n=C_{i}(V_{g}-V_{th})/e$ ($V_{th}$: threshold voltage estimated from Fig. 1(a)).
The hole density is close to the hole density per conducting layer expected at room temperature, $n=1.8\times 10^{14}$ cm$^{-2}$.
At low temperatures, $R_{H}$ appears to be almost independent of temperature (Fig. 2(d)), namely, $\mu _{H}$ rather than $n$ is thermally activated.
This is in contrast with the behavior of $\mu _{\rm FE}$ in the same region, which increases slightly with decreasing temperature.
Note that the decrease in $R_{H}$ at temperatures above 25 K in sample A is attributed to the contribution of thermally activated holes in the bulk.

In $V_{g}$-sweeping, we observed a shift in $R_{H}$ between the bulk and field-induced interface (Fig. 3(a)).
The $R_{H}$ values are almost constant at $V_{g}<0$ and they represent the carrier density in the bulk.
Applying a positive gate voltage reduces $R_{H}$ to a temperature independent value of approximately 4 $\Omega$/T.
According to the classical expression for a two-carrier system, $R_{H}$ is given by
\begin{equation}
R_{H}=\frac{\mu _{h}^{2}n_{h}-\mu _{e}^{2}n_{e}}{e(\mu _{h}n_{h}+\mu _{e}n_{e})^{2}}
\end{equation}
where $\mu _{h}$ ($\mu _{e}$) and $n_{h}$ ($n_{e}$) denote the mobility and density of holes (electrons), respectively.
$\mu _{h}$ and $n_{h}$ are respectively set to the mobility and carrier density of thermally activated bulk holes.
In the case of a conventional FET, $\mu _{e}$ is replaced with the measured $\mu_{\rm FE}$ and $n_{e}$ is estimated from $n_{e}=C_{i}(V_{g}-V_{th})/e$.
$R_{H}$ should be negative at $V_{g}>V_{th}$ and then vary almost proportionally with $1/(V_{g}-V_{th})$ (inset of Fig. 3(a)).
However, no sign reversal was observed between the bulk and field-induced carriers over the entire temperature range. 
The conventional capacitance model, $n_{e}=C_{i}(V_{g}-V_{th})/e$, is not valid for the present system.

In order to understand the effect of the gate voltage in this system, we performed a simple simulation of $R_{H}$ by making the following assumptions.
The first assumption is that electron carriers can be ignored in the system so that $\mu _{e}$ and $n_{e}$ in Eq. (2) can be replaced with the values for the surface hole carriers (the negative sign in the numerator becomes positive for hole carriers).
The second assumption is that the surface hole density is constant ($n=1.6\times 10^{14}$ cm$^{-2}$) so that the mobility is enhanced by increasing $V_{g}$. 
The surface hole mobility exhibits the same $V_{g}$ dependence as the conductivity because $\sigma =en\mu$.
This delocalized hole model reproduces the experimental data well, as shown in the inset of Fig. 3(a).
Thus, the straightforward interpretation is that a positive $V_{g}$ activates the mobility of hole carriers.
The hole density is around $n=1.6\times 10^{14}$ cm$^{-2}$ corresponding to 90\% of the first Brillouin zone in a single conducting layer of $\kappa$-Br.
The effective mobility (but not the concentration) of the field-induced carriers increases with an increase in the gate voltage (Fig. 3(b)).

\begin{figure}[htbp]
 \begin{center}
 \includegraphics{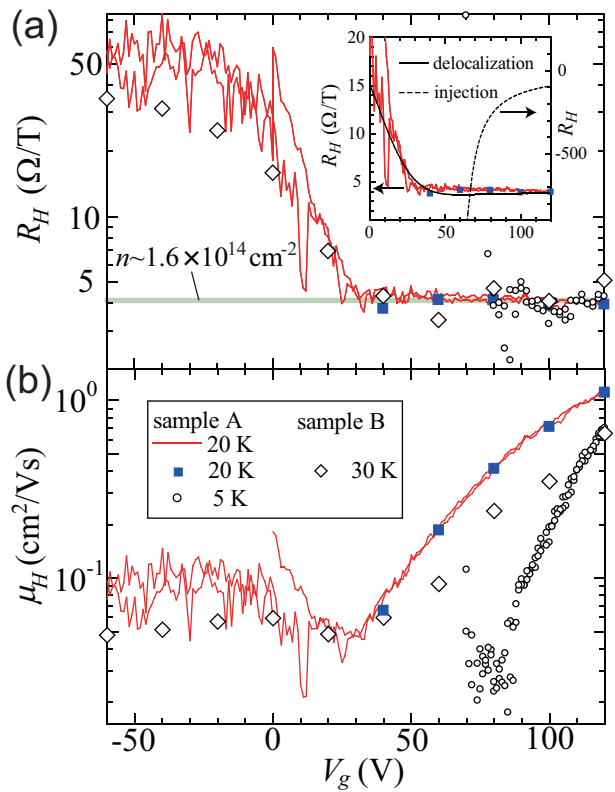}
 \end{center}
 \caption{(a) $V_{g}$ dependence of $R_{H}$. The blue squares and open diamonds are from $B$-sweeping. Other data points are obtained by $V_{g}$-sweeping; both the forward and backward data are shown. The inset shows simulations for electron injection and hole delocalization (see the text) plotted on a linear scale. (b) The $V_{g}$ dependence of $\mu _{H}$.}
 \label{fig:Fig3.eps}
\end{figure}

Below, we summarize the observed characteristics of the field effect.
(i) The conductivity obeys the formula $\sigma _{\Box} = \sigma _{0}\exp(-W/k_{B}T)$, where $\sigma _{0}$ is a constant and $W$ depends on $V_{g}$.
(ii) 1/$eR_{H}$ increases when a positive gate voltage is applied, but its sign remains positive.
(iii) In the $V_{g}$ vs $R_{H}$ plot, $R_{H}$ appears to alternate between the bulk and surface values. The value of 1/$eR_{H}$ at high $V_{g}$ gives a hole density of about $n=1.6\times 10^{14}$ cm$^{-2}$ corresponding to 90\% of the first Brillouin zone in a single conducting layer.
(iv) When the temperature is varied, $R_{H}$ remains almost constant at the gated surface, indicating that the Hall mobility, rather than the concentration, of the field-induced carrier is thermally excited.

The constant $\sigma _{0}$ indicates that the macroscopic phase separation, which is observed in the bandwidth-controlled Mott transition of $\kappa$-Br \cite{Sasaki2}, does not occur by a gate voltage. 
Since increasing the chemical potential by applying a positive gate voltage should inject only electron carriers into a band insulator, (ii) and (iii) indicate that the suppression of the activation energy is due to the collapse of the charge gap in strained $\kappa$-Br. 
In other words, the hole carriers that are initially localized due to the electron correlation become delocalized when a gate voltage is applied. 
These results are consistent with the continuous character of the bandfilling-controlled Mott transition in which the charge mass diverges with approaching half-filling \cite{Furukawa}.

The temperature and gate voltage independence of $R_{H}$ (at high $V_{g}$) shown in (iii) and (iv) implies hopping transport without excitation in the carrier density, conflicting with the mobility edge model proposed for the insulating region of Si-MOSFETs \cite{Ando}. 
In the future, we intend to investigate the transport mechanism when a gate voltage is applied in more detail by performing additional experiments at lower temperatures. 
Furthermore, it will be important to investigate the capacitance and the magnetic properties.

As far as we know, the present experiment is the first continuous observation of $R_{H}$ for a FET structure using a Mott insulator.
The results reveal the clear difference between this device and conventional FET devices, namely that the intrinsic localized carriers have a finite mobility when a gate electric field is applied. 
Electrostatic tuning is a powerful tool for investigating the physics of Mott insulators. Organic $\pi$ electron systems are particularly suitable for this method because of their high purities, low carrier concentrations and simple electronic structures.

\begin{acknowledgments}
We would like to acknowledge Drs. H. Taniguchi, T. Minari, Y. Nishio and K. Kajita for valuable discussions, Dr. K. Kubo for helpful advice on crystal preparation, and Drs. S. Niitaka and H. Katori for assistance with using the instrument. This work was partially supported by Grants-in-Aid for Scientific Research (Nos. 16GS0219 and 20681014) from the Ministry of Education, Culture, Sports, Science and Technology of Japan.
\end{acknowledgments}

\end{document}